\documentclass{ws-ijgmmp}

\begin{document}

\markboth{Tom Lawrence}
{Tangent space symmetries in general relativity and teleparallelism}

%
\catchline{}{}{}{}{}
%

\title{Tangent space symmetries in general relativity and teleparallelism}

\author{Tom Lawrence}

\address{Ronin Institute for Independent Scholarship, 127 Haddon Place\\
	Montclair, New Jersey 07043-2314, United States of America\\
	\email{tom.lawrence@ronininstitute.org}}

\maketitle

\begin{history}
\received{(Day Month Year)}
\revised{(Day Month Year)}
\end{history}

\begin{abstract}
This paper looks at how changes of coordinates on a pseudo-Riemannian manifold induce homogeneous linear transformations on its tangent spaces. We see that a pseudo-orthonormal frame in a given tangent space is the basis for a set of Riemann normal coordinates. A Lorentz subgroup of the general linear transformations preserves this pseudo-orthonormality. We borrow techniques from the methodology of non-linear realizations to analyze this group-subgroup structure.

`Parallel maps' are used to relate tangent space at different points. `Parallelisms' across a finite region of the manifold may be built up from them. These are used to define Weitzenb\"{o}ck connections and Levi-Civita connections.

This provides a new formulation of teleparallel gravity, in which the tetrad field is viewed as a field-valued group element relating the coordinate basis to the frame basis used in defining a parallelism. This formulation separates the metric degrees of freedom from those associated with the choice of parallelism. The group element can be combined by matrix multiplication with Lorentz transformations of frame or with other Jacobian matrices. We show how this facilitates a new understanding of inertial forces and local Lorentz transformations.

The analysis is also applied to translations of the coordinates. If they are constant across spacetime, this has no effect on the tangent space bases. If the translation parameters become fields, they induce general linear transformations of the coordinate basis; however, the tetrad components can only be expressed in terms of translations on a flat spacetime.

\end{abstract}


\section{Introduction \label{Intro}}

We start this paper with some historical background on the main concepts, focusing on how they were first used in, or emerged from, physics research. General relativity (GR) is covered very briefly, as it so well known. It is assumed that readers will have at least some basic familiarity with teleparallelism. This subject is covered to the extent needed to set out the main debates and concepts addressed in this paper. Non-linear realizations are covered in more detail, because few researchers in GR or teleparallel gravity are likely to have a solid understanding of both this subject. (These topics are covered roughly in chronological order, splitting the narrative on teleparallelism either side of that on non-linear realizations.)

This introductory section then concludes by summarizing the main concepts in this paper, the notation used and the structure of the paper.

\subsection{General relativity}

Einstein spent many years developing GR, completing it with Hilbert in 1915. It takes spacetime to be a pseudo-Riemannian manifold. This means that tangent spaces admit a ten-component metric. These ten components are the basic variables in the gravitational sector of the theory. More generally, in an $N$-dimensional spacetime, there would be $\frac{1}{2} N(N+1)$ of these. Neighbouring tangent spaces are compared using the Levi-Civita connection, allowing covariant derivatives to be defined. This connection is uniquely defined on a given spacetime for a given coordinate system and is constructed from the metric - or more precisely, its derivatives and its inverse. Its field strength is the Riemann curvature of the spacetime, which represents the gravitational field. A contraction of this tensor, the Ricci tensor, is non-zero wherever matter is present, and its relation to another tensor describing the matter distribution provides the field equation of the theory.

\subsection{Teleparallelism - the early years}

When GR was developed, the only other fundamental interaction that was recognized by physics was electromagnetism. Einstein was keen to extend the theory to incorporate electromagnetism in a geometric way. He tried a number of different approaches to this\cite{SauerUFT}.

One approach was `Fernparallelismus', translated as `distant parallelism' and now known as `teleparallelism'\cite{SauerTP}. He noted that an $N$-bein field has $N^2$ independent components. The components of the metric can be written as functions of these, but then there are $\frac{1}{2} N(N-1)$ degrees of freedom contained in the $N$-bein field which describe invariances of the metric\cite{Einstein1928}. His idea was that these additional degrees of freedom could be used in describing electromagnetism. In defining an $N$-bein field across the spacetime manifold, he needed to use a new type of connection, which he discovered had already been investigated by Weitzenb\"{o}ck, Cartan and others\cite{SauerTP,Delphenich}.

This approach was unsuccessful in its aim and it appears that papers on the subject were only published sporadically in the decades following the Second World War. While not strictly a paper on teleparallelism, the work of Utiyama\cite{Utiyama} is worth noting in the development of the theory of the `tetrad', as the four-dimensional $N$-bein field has come to be known\footnote{In the usual formulation of teleparallel gravity, there is a pseudo-orthonormal frame associated with each point in spacetime. Each `leg' of this vielbein or tetrad can be decomposed in the coordinate basis, giving rise to a 16-component matrix, which carries the field degrees of freedom. The frame basis and the 16-component matrix are both sometimes referred to as the `tetrad', as the 16 fields of the latter are seen as components of the former. In this paper, they are viewed as fundamentally separate objects: the matrix is an element of the general linear group, while the 	frame basis is an element of its carrier space. We only use the word `tetrad' to make contact with the usual formulation, and when we use it, we mean the field-valued matrix.}. His aim was to provide a common gauge theory description of gravity, electromagnetism and the theory of Yang and Mills\cite{YM} which had been published the previous year. Utiyama considered gravity as a gauge theory of the Lorentz group, although he saw general relativity as also having a separate symmetry under coordinate transformations. 

This view of these symmetries as separate has persisted until today, but we show in this paper that this is not actually the case. Lorentz transformations form a subgroup of the general linear transformations induced by changes of coordinates. Teleparallel papers of the 1960s and 1970s seem to regard them as separate transformations for the purpose of calculations (see \cite{PP,Moeller} and references therein), although Pellegrini and Plebanski\cite{PP} talk about ``the group of coordinate transformations'' and its ``subgroup of constant Lorentz rotations of the tetrad''.

\subsection{Coset space decompositions and non-linear realizations}

Non-linear realizations became an active area of research in the 1960s. This started with the non-linear sigma model, introduced by Gell-Mann and L\'{e}vy in 1960\cite{G-ML}. In this model, a four-component multiplet of Lorentz scalar fields was constrained by fixing the norm of the multiplet. This allowed them to eliminate one component from the Lagrangian. This elimination left the Lagrangian depending non-linearly on the remaining components. (The resulting field space was diffeomorphic to $S^3$, parametrized by the three remaining field components, although this was not stated in the paper.) 

This was followed by many papers in which the Lagrangian was invariant under an internal symmetry group - typically a chiral group - but was constructed non-linearly from the fields on which the transformations act. In 1968, Coleman \emph{et al}\cite{CWZ} identified a common feature of this research as crucial to further analysis: that the fields transformed linearly only under a subgroup of these symmetries and transformed in a non-linear way under the remainder of the group. (See references within \cite{CWZ} for the earlier papers.) 

This paper showed how it was possible to classify all such non-linear realizations, presenting a comprehensive analysis based on group theory. A companion paper provided the general method for constructing Lagrangians with the required symmetries, based on covariant derivatives\cite{CCWZ}. The following year, a paper by Salam and Strathdee\cite{SS1} showed how this theory was applicable whenever a Lie group symmetry was spontaneously broken to a continuous subgroup\footnote{This was also shown in the specific case of chiral SU(3) by Honerkamp\cite{Honerkamp}.}. (Such spontaneous symmetry breaking in field theory was first proposed by Goldstone\cite{Goldstone} in 1960 - the same year as the Gell-Mann and L\'{e}vy paper, and indeed, in the same journal.)

This body of theory has been developed further since 1969, notably in Meetz\cite{Meetz}, Isham\cite{Isham-met,Isham-emb1}, Balachandran \emph{et al}\cite{BST} and Boulware and Brown\cite{BB}. However, the necessary background for the analysis in the remainder of this paper can all be found in Coleman \emph{et al}\cite{CWZ} and Salam and Strathdee\cite{SS1}. 

The starting point of Coleman \emph{et al} was to note that a subgroup $H$ of a group $G$ could be used to partition $G$ into cosets of the form $gH$, where $g \in G$. Now if $G$ and $H$ are both linear Lie groups, we may write any element of $G$ in the form of an exponential of a matrix in its Lie algebra, and similarly for any element of $H$. However, if $h \in H$, then the cosets $gH$ and $ghH$ are identical. It is then easy to see that we can uniquely write $gH$ in the form
\begin{equation} \label{gH}
g H = \mathrm{e}^{\mathrm{i} \theta^a M_a} \left\lbrace \mathrm{e}^{\mathrm{i} \theta^A M_A} \; \forall \, \theta^A \right\rbrace 
\end{equation}
where $M_A$ are the generators of $H$, $M_a$ are the remaining generators of $G$ (now called the `broken generators'), the $\theta$ are group parameters associated with these various generators and repeated indices are summed over. The first factor here is a representative of the coset, denoted $L$:
\begin{equation}
L = \mathrm{e}^{\mathrm{i} \theta^a M_a} \, .
\end{equation}
Then, because each element of $G$ is in only one coset, it can be uniquely decomposed in the form
\begin{equation} \label{g-decomp}
g = L h \, .
\end{equation}

A Lorentz scalar multiplet of $G$ can be constrained either `by hand' (as in the non-linear sigma model) or through a potential with degenerate minima constructed from the multiplet (as in the Goldstone mechanism). The vacuum manifold/constrained field space is then diffeomorphic to $G/H$, where $H$ is the subgroup under which a point in this field space is invariant. This field space is then parametrized by the coset space parameters $\theta^a$ - that is, there is one `Goldstone boson' for each of these parameters.

Using the parametrization of the coset space (\ref{gH}), Coleman \emph{et al} analyzed the transformation of these fields. If we start with a coset $L_0 H$, the action of another element $g$ on it is as follows:
\begin{equation} \label{coset-map}
g: L_0 H = \mathrm{e}^{\mathrm{i} \theta^a M_a} \left\lbrace \mathrm{e}^{\mathrm{i} \theta^A M_A} \; \forall \, \theta^A \right\rbrace  \mapsto L' H = \mathrm{e}^{\mathrm{i} \theta'^a M_a} \left\lbrace \mathrm{e}^{\mathrm{i} \theta^A M_A} \; \forall \, \theta^A \right\rbrace \, .
\end{equation}
Thus $L_0$ is mapped to a new coset space representative
\begin{equation}
L' = \mathrm{e}^{\mathrm{i} \theta'^a (\theta^a, g) M_a} \, .
\end{equation}
In particular, under the action of an element of $H$, they showed that
\begin{equation}
h: L_0 \mapsto L' = h L_0 h^{-1} \, .
\end{equation}

They also showed that $L^{-1}$ can be used to reduce any other multiplet of $G$ in the system to a multiplet of $H$ only, but we do not reproduce this analysis here because it will not be needed in this paper.

The coset space representative $L$ also turned out to be crucial to constructing covariant derivatives. From it, one can construct a derivative based on the Maurer-Cartan form, $L^{-1} \partial_\mu L$. This takes values in the algebra of $G$. It has a Cartan decomposition into the subgroup and coset space parts of the Lie algebra:
\begin{equation}
L^{-1} \partial_\mu L = 2 \mathrm{i} (v^A_\mu M_A + a^a_\mu M_a) \, .
\end{equation}
It turns out that the $a^a_\mu$ are the covariant derivatives for the Goldstone bosons:
\begin{equation}
D_\mu M^a = a^a_\mu \, ,
\end{equation}
while the $v^A_\mu$ are the parameters of the connections involved in the covariant derivatives of any other fields in the system:
\begin{equation}
D_\mu \psi = \partial_\mu \psi + \mathrm{i} v^A_\mu S(M_A) \psi \, ,
\end{equation}
where $S(M_A)$ is the representation of the subgroup generators appropriate to the multiplet $\psi$.

In later sections of this paper, we apply some of the above theory to tangent space symmetries. It should be noted that this is far from the first time that this theory has been applied to spacetime symmetries. Indeed, Salam and Strathdee provided a companion paper to that mentioned above, in which $G$ was the conformal group and $H$ was its Lorentz subgroup\cite{SS2}. Since then, there have been many others (see, for example, \cite{Hankey,DT,TM,Kirsch}). Indeed, in recent decades, there have even been papers on non-linear realizations in teleparallel gravity\cite{L-PTT,TT1,TT2} and symmetric teleparallel gravity\cite{Koivisto}. Many of these papers have had an explicit aim of establishing a non-linearly realized gauge symmetry, often as a route to quantization. The analysis in the current paper is quite different. We merely observe the presence of a group-subgroup structure, in which the metric degrees of freedom are associated with the coset space parameters, and borrow techniques from the theory of non-linear realizations for analysing it. There is actually a non-linear realization hidden within this structure, but it is not utilized in the analysis in this paper.

\subsection{Teleparallelism - recent decades}

Since the late 1970s, teleparallelism has undergone a revolution. This has resulted in the Teleparallel Equivalent of General Relativity (TEGR) and several families of `modified' theories. Details of their genesis can be found in a 2013 review by Maluf\cite{Maluf}. All of them use the Weitzenb\"{o}ck connection, which has zero field strength but does have a torsion. 

It is a scalar constructed from this torsion which is used in the action of TEGR, in place of the Einstein-Hilbert action. The action is varied with respect to the tetrad and results in the same field equations as GR. 

In the modified theories, the Lagrangian contains torsion scalars which are more general than that used in TEGR. It is now known the universe is expanding and that the rate of this expansion is accelerating. Cosmological observations are becoming increasingly precise, revealing growing detail about this expansion and about the early epochs of the universe. These do not fit easily within GR as it stands and researchers are looking for consistent teleparallel theories which fit the observations. 

Teleparallel theories of gravity are also seen as having other advantages over GR. 

Firstly, they engage directly with spacetime symmetries, as is explored in this paper. The connections contained in covariant derivatives therefore appear to be gauge fields for these symmetries. The other fundamental interactions are also described in terms of gauged symmetries, and quantized on that basis. These similarities have provided a basis for working towards a quantum field theory for gravity.

There is some debate, however, as to which symmetries are gauged in teleparallel theory. Hayashi and Nakano\cite{HN}, for example, took a different view from Utiyama (see above), publishing a paper in 1978 which proposed that gravity can be expressed as a gauge theory of translations. This was based on the construction of a derivative operator which is covariant under local translations. From this point onwards, it has often been stated that teleparallel gravity constitutes a gauge theory of translations. Recent publications to this effect by Aldrovandi and Pereira\cite{AP} and Pereira and Obukhov \cite{PO} have been disputed by Fontanini \emph{et al}\cite{FHLD1,FHLD2}. They point to the role of the general linear group as the structure group of the tangent bundle. Hohmann \emph{et al}\cite{KHZ}, meanwhile, do not view general linear transformations and translations as separate, stating that ``A displacement transformation can be seen, from the passive perspective, as a general coordinate transformation. Obviously, such a transformation is included in the GL (General Linear) transformations. In a GL gauge connection, the displacement component is precisely that which has neither curvature nor torsion''.

In this paper, we look at translations mainly as changes of coordinates. It is shown that such passive translations do not act directly on the tangent bundle. Global translations do not act on the tangent spaces at all. The action on the tangent space induced by local translations is simply that of the general linear group, albeit represented as a displacement of the basis rather than a contraction with the basis. This is a subtle advance on the statement made by Hohmann \emph{et al} - we find that \emph{any} general linear transformation on a specific tangent space can be expressed in terms of translations. However, expressing a tetrad \emph{field} in terms of a local translation does constrain its derivatives. Consequently, the Weitzenb\"{o}ck connection can only be constructed from translation parameters in a flat region of spacetime. Nonetheless, the relationship between local translations and the general linear group allows one to write, for example, the connection along a geodesic in terms of the derivatives of the translation parameters.

The displacement of the basis provides a minimal coupling to the derivatives of the translation parameters, in line with the covariant derivative operator of Hayashi and Nakano\cite{HN}. However, an inhomogeneous displacement of the vector components cannot be induced by a coordinate transformation. Allowing for such a transformation would constitute extending general relativity. 

This would all seem to support many of the arguments advanced by Fontanini \emph{et al}. However, it may not preclude teleparallel gravity being a gauge theory of translations expressed as \emph{point transformations}. We touch on this only briefly at the end of this paper and note that it is an area worthy of further study.

Secondly, these theories are particularly useful for calculations concerning conservation of energy density, due to the way in which the energy-momentum density of the gravitation field may be separated out from inertial effects. Deep and informative analysis on this issue was carried out by Aldrovandi \emph{et al}\cite{ALP}. 

While most of their analysis seems valid, it suggests that when the teleparallel spin connection has a non-zero value, this is associated purely with inertial effects. This is a claim often found in the literature. For example, Kr\u{s}\u{s}ak and Saridakis\cite{KS} state that ``various spin connections represent different inertial effects for the observer'', while Golovnev \emph{et al}\cite{GKS} state that ``the spin connection is then associated solely with inertia''. We show in Sections \ref{chart} and \ref{inertial} that this is not in fact the case. The spin connection depends on the choice of frame \emph{used to define the parallelism}. Inertial effects, by contrast, are due to the choice of local inertial frame for the observer, which is entirely determined by the coordinate system.

\subsection{Overview and structure of this paper \label{OandS}}

This paper looks at the way in which changes of coordinates on a pseudo-Riemannian manifold form an infinite dimensional group, and induce homogeneous linear transformations of the coordinate basis on each tangent space. This can be seen as a homomorphism from the covariance group to the structure group of the tangent bundle, a general linear group. Pseudo-orthonomality is preserved under a Lorentz subgroup of this general linear group.

Recently, there has been considerable debate in the teleparallel gravity research community about this Lorentz subgroup of symmetries - its physical meaning and whether the degrees of freedom in the Lorentz connection represent physical fields\cite{KS,GKS,BCKPVDH,BFFG}. A summary of recent papers and the current state of understanding can be found in a paper by Bejarano \emph{et al}\cite{BFFG}). 

The approach of this paper is to note that the $N^2$ degrees of freedom identified by Einstein - in our case 16 - are those of the general linear group. We then separate them into those which are contained in the metric and those which are associated with the Lorentz subgroup using a decomposition analogous to (\ref{g-decomp}). In outline, we will do this as follows.

We denote the general linear group $J$ and its Lorentz subgroup $I$. An arbitrary element $j$ can then be decomposed as
\begin{equation}
j = l i \, ,
\end{equation}
where $i \in I$ and $l$ is a representative of the coset space $J/I$. Thus if we write the generators of $I$ as $M_{\mu \nu}$, the remaining generators of $J$ as $m_{\mu \nu}$, and their corresponding parameters $\Omega^{\mu \nu}$ and $\omega^{\mu \nu}$ respectively, we have
\begin{equation} \label{l-omega}
l = \mathrm{e}^{\mathrm{i} \omega^{\mu \nu} m_{\mu \nu}}
\end{equation}
and
\begin{equation} \label{i-Omega}
i = \mathrm{e}^{\mathrm{i} \Omega^{\mu \nu} M_{\mu \nu}} \, .
\end{equation}
The group $J$ is used to carry out changes of basis on the tangent space. Its elements are matrices which act on a basis by contraction to map it to another basis. This includes (pseudo-orthonormal) frame bases (which, as we will see, can be considered locally as coordinate bases for a set of Riemann normal coordinates). Thus in a specific tangent space, the transformation $j_0$ between a given orthonormal frame and a given coordinate basis can be decomposed as
\begin{equation} \label{j0-decomp1}
j_0 = l_0 i_0 \, ,
\end{equation}
with $l_0$ and $i_0$ taking the forms (\ref{l-omega}) and (\ref{i-Omega}) above. 

To construct covariant derivatives, we need a coordinate-independent way of comparing vectors in different tangent spaces. We do this by choosing a `parallel map' from one tangent space into a second, which is linear over the space and preserves pseudo-orthonormality. This defines a frame in the second space as the image of the frame in the first. By constructing such a map to every tangent space in a coordinate neighbourhood, we build up a `parallelism', with the frame bases forming a frame field. If the frame field is continuous over the neighbourhood, the connection associated with this parallelism is a Weitzenb\"{o}ck connection.

Any continuous frame field can be used to construct a Weitzenb\"{o}ck connection. They are all related by local Lorentz transformations. This is a crucial difference between Levi-Civita and Weitzenb\"{o}ck connections. To calculate the functional form of a Levi-Civita connection on a given spacetime, one needs only to choose a coordinate system. For a Weitzenb\"{o}ck connection, by contrast, one must also choose a parallelism.

Once a coordinate system and a parallelism have been chosen, this specifies the coordinate basis and frame basis at every point. (\ref{j0-decomp1}) can then be used to decompose the transformation between them, with $j_0, l_0, i_0$ promoted to fields.

We will denote the coordinate basis $\mathbf{e}_\mu$ and the frame basis associated with the parallelism $\hat{\mathbf{n}}_\mu$, with the hat denoting orthonormality, so that
\begin{equation}
\mathbf{e}_\mu = (j_0)_\mu{}^\xi \hat{\mathbf{n}}_\xi \, .
\end{equation}
While it is possible to identify these bases with differential operators, as is done in the formalism of differential forms, this is not necessary for the analysis in this paper. We stick to the more abstract form above, as this better represents the bases as belonging to the carrier space of $J$. Crucially, $\mathbf{e}_\mu$ and $\hat{\mathbf{n}}_\mu$ belong to the \emph{same} carrier space - they are bases on the same tangent space, and for this reason they carry the same index. This is a major difference from the usual tetrad formulation of teleparallel gravity, in which the frame basis usually carries a Latin index and the coordinate basis a Greek one. This means that $j_0, l_0, i_0$ have two indices of the same type, as appropriate to group elements, and they can be combined with other group elements by matrix multiplication. This facilitates changing coordinates or changing paralellism and is what gives this formulation its power. We will explain the relation between this formulation and the usual tetrad one shortly in this subsection.

As we will see, the decomposition (\ref{j0-decomp1}) allows us to define an `intermediate' basis, $\hat{\mathbf{k}}_\xi$. The resulting situation can then be schematically represented as in Fig. 1.

\begin{figure} [ht]
	\centering
	\includegraphics[width=1\linewidth]{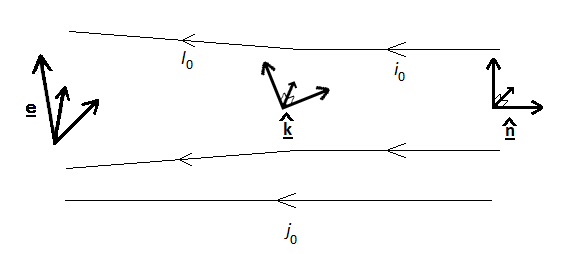}
	\caption{Action of $j_0$}
\end{figure}

A change of coordinates induces an action $j(u)$ on $\mathbf{e}_\mu$ and therefore acts on $j_0$ from the left. This will change both the $l_0$ and $i_0$ factors. A change of parallelism is achieved by applying $i(u)$ to $\hat{\mathbf{n}}_\mu$ and therefore acts on $j_0$ from the right. This changes just the $i_0$ factor.

This decomposition is helpful when working with the Weitzenb\"{o}ck connection, because, as we shall see, its associated Lorentz connection depends only upon $i_0$ and its derivatives and hence upon the $\Omega^{\mu \nu}$ fields. Similarly, the metric is a function of $l_0$ and hence the $\omega^{\mu \nu}$ fields only. These will transform as a non-linear realization of $J$. However, in practice, it is easier to work with $i_0$ and $l_0$ than $\Omega^{\mu \nu}$ and $\omega^{\mu \nu}$ and these group parameters are not needed at all in the analysis in this paper. The expressions (\ref{l-omega}) and (\ref{i-Omega}) are implicitly assumed throughout this paper, but these forms are never explicitly used. (They could be important, though, when trying to identify explicit forms for the `tetrad fields' in particular situations, as in \cite{HJKP}.)

This deliberate separation of these degrees of freedom at the outset distinguishes this paper from the existing literature on teleparallelism. As noted by Golovnev \emph{et al}, until recently, most research has been carried out in the `Weitzenb\"{o}ck gauge', in which the Weitzenb\"{o}ck spin connection vanishes. This happens when the frame field used to define the Weitzenb\"{o}ck connection\footnote{The Weitzenb\"{o}ck connection is described in qualitative terms above and defined more precisely in Section \ref{chart}. Some authors, such as Bejarano \emph{et al}\cite{BFFG}, use the phrase `Weitzenb\"{o}ck connection' to mean this connection specifically in the Weitzenb\"{o}ck gauge. This has probably arisen from the practice of working in this gauge by default. By contrast, we use it in the wider sense given by Pereira\cite{Pereira}.} is such that $i_0$ is constant (a global Lorentz transformation). These are referred to as `proper frames' by Kr\u{s}\u{s}ak and Saridakis, while Golovnev \emph{et al} call this a `pure-tetrad approach'. (`Proper frames' therefore differ from $\hat{\mathbf{k}}_\xi$ by at most a global Lorentz transformation.) Golovnev \emph{et al} comment that ``Often the pure-tetrad approach is assumed explicitly, often it is not clear what is assumed'. 

However, in recent years, it has been realized that working in the Weitzenb\"{o}ck gauge violates Lorentz symmetry, which is seen as undesirable. Some authors have tried to remedy this (for example  \cite{KS,GKS,BCKPVDH}), as described by Bejarano \emph{et al}\cite{BFFG}, by taking the theories in the Weitzenb\"{o}ck gauge and supplementing them with a non-zero Lorentz connection, to `covariantize' them. These have started to reveal some of the structure described in this paper. However, much of this research has been hampered by two things. The first is not understanding the Lorentz group as a subgroup of the structure group on the tangent bundle. The second is confusion about its relation to inertial effects - which can only be understood when the group theory aspects are clear. Consequently, these papers have adopted a wide variety of concepts, terminology and symbols, which are inconsistent across papers and sometimes within papers. 

For example, Bejarano \emph{et al} use $\mathbf{e}_a$ for a generic frame basis, with components $e_a^\mu$ (when the coordinate basis is $\partial_\mu$). Golovnev \emph{et al} use $e_a^\mu$ for a `proper tetrad'. B\"{o}hmer \emph{et al}\cite{BCKPVDH}, on the other hand, reserve $e_a$ for a frame basis on Minkowski (flat) spacetime, specializing further to $e'_a$ for a holonomic frame on this spacetime, and use $h^a{}_\mu$ for a `non-trivial tetrad field'. Kr\u{s}\u{s}ak and Saridakis\cite{KS} use $h^a{}_\mu$ for a generic tetrad, but also introduce a `reference tetrad' $h^a_{(r)}{}_\mu$, based on the limit $G \rightarrow 0$. Most authors simply reverse the positions of indices to denote the inverse tetrad, but some use a different symbol - for example Hohmann \emph{et al}\cite{HJKP} use $\theta^a{}_\mu$ for the tetrad and $e_a{}^\mu$ for the inverse tetrad. 

Given the inconsistency of notation and its unsuitability for the formulation of teleparallel gravity contained in this paper, I have decided to stick to the notation I have used for many years prior to learning of these debates. The notation adopted in this paper emphasizes, and is appropriate to, a group theory approach. Inverse group elements, for example, are denoted with a suffix of $-1$ rather than just changing the position of the indicies. (Another example of this difference is that in this paper, a Lorentz factor $i_0$ is included in the Weitzenb\"{o}ck spin connection from the start; any further Lorentz transformation of the frame is then denoted $i$. In the existing literature, by contrast, $\Lambda^a{}_b$ is often used for any Lorentz transformation; if it is used in the Weitzenb\"{o}ck spin connection it may differ from the $i_0$ here by a global factor.)

However, this paper does follow the notation of Pereira\cite{Pereira} in using a dot above a connection or covariant derivative to specify that it is a Weitzenb\"{o}ck connection or covariant derivative. Similarly, circles above connections or covariant derivatives denote that they are Levi-Civita ones.

Indices are usually used explicitly, except where their suppression makes a point of group theory clearer. Greek letters are used for indices, but we avoid using the letter $\tau$ for this purpose - this is exclusively used to denote proper time (and on occasion is used as a subscript in this context).

Where it is necessary to specify which coordinate system a set of tensor components relates to, this will be done by putting it in brackets in a superscript or subscript. For example, the components of a vector $\mathbf{V}$ in a coordinate system $u'^\mu$ will be written $V^\mu_{(u')}$.

In this analysis, it is very easy to confuse the value of a function at a point with its functional form. Consequently, we will often state explicitly which point a function is evaluated at, if this is what we are doing.

The language of fibre bundles is largely avoided, as the approach is intended to emphasize firstly how the group transformations are induced by changes of coordinate, and secondly the physical interpretation of these transformations.

The theory in this paper assumes a pseudo-Riemannian spacetime, in which observations are made by a classical (point-like) observer. Measurements of field configurations taking values in a Hilbert space are explicitly excluded from the analysis. It proceeds from this point with geometric arguments, which are independent of the action being used. While it is phrased in terms of a four-dimensional spacetime, the theory is also valid for a higher-dimensional spacetime. 

While the theory presented above summarizes much of the content of Sections \ref{TP} to \ref{chart}, there are many subtleties and traps for the unwary in this analysis. We therefore take the arguments step-by-step from first principles. It may seem unnecessarily slow and basic to some readers, but this is what has led to the insights in this paper into the current debates, and it seems prudent to err on the side of caution to avoid further misunderstandings.

Section \ref{TP} looks at the covariance group of coordinate transformations. It finds the induced action of this group on the coordinate bases and vector components in a tangent space \emph{at a single point}. The Jacobian matrix is seen to be an element of a general linear group. (This analysis will be crucial when looking at translations in Section \ref{Transl}.) The metric is then defined. The pseudo-orthogonal invariance group of the Minkowski metric is used to partition the general linear group into cosets and this leads to a natural decomposition of the change of frame. 

This analysis is then extended to a curve in Section \ref{curve}, allowing one to define connections and covariant derivatives. The transformation properties of these are identified. It is then extended further to a four-dimensional chart on the spacetime in Section \ref{chart}. Choices of connection and their associated covariant derivative are then available and we look at the features of these (in particular, the Weitzenb\"{o}ck and Levi-Civita connections) and the relationships between them. In particular, we look at what happens on a geodesic and how Riemann normal coordinates may be applied, as well as field strengths of the connections and their associated Lorentz connections. We consider two different decompositions of the Weitzenb\"{o}ck connection: that using the coset space decomposition and that using its contorsion.

Sections \ref{TP} to \ref{chart} therefore work carefully through the geometric and group theoretic conceptual structure of the analysis. In Section \ref{inertial}, we turn to how this can be used to advance our understanding of physics, focusing on the experience of a classical point-like observer. We start by correcting a misunderstanding that has, quite reasonably, appeared in the literature. We show that, perhaps surprisingly, inertial forces such as centrifugal forces cannot be represented as local Lorentz transformations. (This does not require any of the mathematical framework of the preceding sections - the demonstration is a simple calculation on a flat spacetime, which is then explained in the context of the geodesic equation of GR.) We then proceed to the main topic of this section, the physical interpretation of the Lorentz gauge freedom in teleparallel theories. We show that the choice of parallelism has no effect on the experience of an observer, but this choice is affected by changes of coordinate, which do also affect the experience of the observer. The Lorentz gauge freedom is just a mathematical representation of the freedom to choose the parallelism used to define the Weitzenb\"{o}ck connection.

In Section \ref{Transl} we go back to the covariance group and focus on coordinate transformations in its Poincar\'{e} subgroup - in particular, translations of the coordinates. We look at the map induced on the bases by translations, firstly by ones with constant parameters and then by ones with spacetime-dependent parameters. We look at how these relate to the Jacobian matrices. We invert this to find the transformations of the vector components. We see that inhomogeneous transformations of the vector field components cannot result from coordinate transformations. However, they can result from translations as point transformations, as we show in an appendix to the paper.

Finally, the paper is summarized in Section \ref{Concs}.

\section{The tangent space at a point \label{TP}}

Following GR, we assume spacetime to be a pseudo-Riemannian manifold. We take Nakahara's definition\cite{Nakahara}, which can be summarized as follows. A manifold is a topological space which can be covered by open sets known as `coordinate neighbourhoods'. Each coordinate neighbourhood has a mapping onto $\mathbb{R}^N$ that provides coordinates for the neighbourhood, and where they overlap, the relations between the two sets of coordinates are infinitely differentiable - we shall go slightly further and assume they are analytic. A pseudo-Riemannian manifold is simply one which admits a pseudo-Riemannian metric.

If $u^\xi$ and $u'^\xi$ are two sets of curvilinear coordinates on a coordinate neighbourhood $\Omega$ of a manifold $\mathcal{M}$, the analyticity assumption tells us that we may write
\begin{equation} \label{ct}
u'^\xi = f^\xi + f^\xi{}_\rho u^\rho + f^\xi{}_{\rho \kappa} u^\rho u^\kappa + \ldots
\end{equation}
where the coefficients $f^\xi{}_{\rho \kappa \ldots}$ are real and symmetric on their lower indices and independent of the coordinates.

This can be viewed as a map $a$:
\begin{equation}
a: u^\xi \mapsto u'^\xi \, .
\end{equation}
This can be combined with a map $b$ to a third coordinate system:
\begin{equation}
b: u'^\xi \mapsto u''^\xi \, ,
\end{equation}
so that 
\begin{equation}
ba: u^\xi \mapsto u''^\xi \, .
\end{equation}
Each such map is assumed to be invertible. It is possible to combine three such maps and we easily see that
\begin{equation}
c(ba) = (cb)a \, .
\end{equation}
With the rule for combining these maps, the set of all such maps forms a group, sometimes called the covariance group, which is known to be isomorphic to the diffeomorphism group $\mathrm{Diff}(\mathcal{M})$. From (\ref{ct}), it is clear this is an infinite-dimensional group, with the parameters $f^\xi, f^\xi{}_\rho, f^\xi{}_{\rho \kappa}, \ldots$.

General relativity and teleparallel theories of gravity are constructed to be generally covariant, meaning that equations can be expressed in forms independent of the precise coordinate system being used. This is achievable despite the complexity of the relationship between $u^\xi$ and $u'^\xi$, because general relativity is expressed in terms of tangent vectors, tensors and connections. The fact that the spacetime is a manifold admitting a metric means that it approximates to flat spacetime at each point. This allows one to define a tangent space at each point, the elements of which are vectors. By taking outer products of the tangent spaces and their duals, one can define tensors of higher rank. These have much simpler transformation laws than the underlying coordinates do.

The vectors tangent to the curves of increasing $u^0, u^1$, $u^2, u^3$ at a point $A$ form a basis for the tangent space $T_A \mathcal{M}$, denoted $\mathbf{e}_\mu |_A$ - the `coordinate basis' for $u^\mu$. The \emph{value} of a vector field at $A$ may then be written as a linear sum of this coordinate basis. Using $|_A$ to denote the value of a quantity at $A$, we have
\begin{equation}
\mathbf{V} |_A \in T_A \mathcal{M} = V_{(u)}^\mu |_A \mathbf{e}_\mu |_A  \, .
\end{equation}
Indeed, it can be written as a linear sum of any set of four independent vectors in the space. In particular, it may be written as a linear sum of a second coordinate basis:
\begin{equation}
\mathbf{V} |_A = V_{(u)}^\mu |_A \mathbf{e}_\mu |_A = V_{(u')}^\nu |_A \mathbf{e'}_\nu |_A  \, .
\end{equation}
We find the relations between the two bases by considering two neighbouring points, $A$ and $B$. If they are separated by an infinitesimal interval, the displacement is a vector in $T_A \mathcal{M}$. This may be written in the two coordinate systems as 
\begin{equation} \label{dus}
d u^\mu |_A \, \mathbf{e}_\mu |_A = d u'^\nu |_A \, \mathbf{e'}_\nu |_A \, .
\end{equation}
Now $B$ has coordinates $u^\xi = d u^\xi$ and $u'^\xi + d u'^\xi$. From (\ref{ct}), these are related by
\begin{equation}
u'^\xi + d u'^\xi = f^\xi + f^\xi{}_\rho (u^\rho + d u^\rho) + f^\xi{}_{\rho \kappa} (u^\rho + d u^\rho) (u^\kappa + d u^\kappa) + \ldots
\end{equation}
so 
\begin{equation}
d u'^\xi = f^\xi{}_\rho d u^\rho + f^\xi{}_{\rho \kappa} u^\rho d u^\kappa + f^\xi{}_{\rho \kappa} d u^\rho u^\kappa + \ldots \, ,
\end{equation}
while we can find from first principles that
\begin{equation} \label{Jac-an}
\left. \frac{\partial u'^\xi}{\partial u^\rho} \right|_A = f^\xi{}_\rho + f^\xi{}_{\rho \kappa} u^\kappa + f^\xi{}_{\kappa \rho} u^\kappa + \ldots \, .
\end{equation}
Comparing these last two equations, we find, unsurprisingly, that
\begin{equation}
d u'^\xi = \left. \frac{\partial u'^\xi}{\partial u^\rho} \right|_A d u^\rho \, .
\end{equation}
Substituting this into (\ref{dus}) gives us
\begin{equation}
d u^\mu |_A \, \mathbf{e}_\mu |_A = d u^\mu |_A  \left. \frac{\partial u'^\nu}{\partial u^\mu} \right|_A \mathbf{e'}_\nu |_A \, .
\end{equation}
This same transformation law is valid for any vector:
\begin{equation}
V_{(u)}^\mu |_A \mathbf{e}_\mu |_A = V_{(u)}^\mu |_A \left. \frac{\partial u'^\nu}{\partial u^\mu} \right|_A \mathbf{e'}_\nu |_A \, .
\end{equation}
We can see this as a transformation of either the basis:
\begin{equation} \label{e-to-e'}
\mathbf{e}_\mu |_A = \left. \frac{\partial u'^\nu}{\partial u^\mu} \right|_A \mathbf{e'}_\nu |_A
\end{equation}
or the components:
\begin{equation} \label{V-to-V'}
V_{(u')}^\mu |_A = V_{(u)}^\mu |_A \left. \frac{\partial u'^\nu}{\partial u^\mu} \right|_A \, .
\end{equation}
Thus while the coordinate transformation (\ref{ct}) and the expression for the Jacobian matrix are (possibly infinite) series of polynomial terms, the actual rules for transforming bases (\ref{e-to-e'}) and vector components (\ref{V-to-V'}) are simple homogeneous linear equations. This simplification is arguably the greatest advantage of working with Riemannian or pseudo-Riemannian manifolds. It should be stressed here that the symmetries we are concerned with in this paper are those under \emph{coordinate transformations}. These relate to different ways of breaking the \emph{value} of a vector into components, associated with coordinate systems. Nonetheless, each such component is itself a field, and these may satisfy symmetries under \emph{point transformations} - we return to this subject in Section \ref{Transl}.

The Jacobian matrices for transforming between bases \emph{at $A$} are invertible $4 \times 4$ real matrices. These are matrices of \emph{values}, not \emph{functions} - the matrix in (\ref{e-to-e'}) can be viewed as taking a `slice' of the derivative matrix at $A$. The set of all such matrices thus forms a group $J_A$ which is isomorphic to $GL(4,\mathbb{R})$. 

Two sequential coordinate transformations result in a product of Jacobian matrices, which is the group rule for the group $J_A$:
\begin{equation}
\left. \frac{\partial u''^\nu}{\partial u'^\mu} \right|_A \left. \frac{\partial u'^\mu}{\partial u^\lambda} \right|_A = \left. \frac{\partial u''^\nu}{\partial u^\lambda} \right|_A \, .
\end{equation}
The induced action of the coordinate transformations on the tangent space at $A$ therefore constitutes a homomorphic mapping of the covariance group of $\mathcal{M}$ into $J_A$, with the map given by (\ref{Jac-an}). Note that this map is spacetime-dependent.

This being a pseudo-Riemannian manifold, we can also define a symmetric inner product for each tangent space:
\begin{equation}
(\mathbf{V}, \mathbf{W})_A = (\mathbf{W}, \mathbf{V})_A \in \mathbb{R} \, .
\end{equation}
The image of this map on the coordinate basis is the metric at $A$:
\begin{equation}
g_{\mu \nu} |_A = (\mathbf{e}_\mu, \mathbf{e}_\nu)_A \, .
\end{equation}
and the inner product acts linearly over the tangent space. We can use this to find the transformation of the metric under a change of coordinates.

We can always define a set of coordinates $x^\xi$ for which the basis is pseudo-orthonormal at our chosen point (with respect to the inner product). We will call this `frame basis' $\hat{\mathbf{n}}_\xi$:
\begin{equation} \label{9b}
(\hat{\mathbf{n}}_\xi, \hat{\mathbf{n}}_\rho)_A = \eta_{\xi \rho} \, .
\end{equation}
We will denote the transformation between the chosen frame basis and the chosen (unprimed) coordinate basis $j_0 |_A$:
\begin{equation} \label{j0-action}
(j_0)_\mu{}^\xi |_A = \left. \frac{\partial x^\xi}{\partial u^\mu} \right|_A \in J_A: \hat{\mathbf{n}}_\mu \mapsto \mathbf{e}_\mu = (j_0)_\mu{}^\xi  \hat{\mathbf{n}}_\xi \, ,
\end{equation}
while $j |_A$ will be used for a generic change of basis - for example,
\begin{equation} \label{basis-change}
j |_A \in J_A : \mathbf{e}_\mu \mapsto \mathbf{e'}_\mu = j_\mu{}^\nu \mathbf{e}_\nu \, .
\end{equation}
Note that in this formalism, $V^\mu$ consequently transforms according to:
\begin{equation}
j |_A: V^\mu_{(u)} |_A \mapsto V^\mu_{(u')} |_A = V^\nu_{(u)} |_A (j^{-1})_\nu{}^\mu  \, .
\end{equation}

As mentioned in the introduction, $j_0$ can be decomposed using a pseudo-orthogonal subgroup. The Minkowski metric\footnote{It has emerged that some authors refer to this as the `Cartesian Minkowski metric' - we follow the terminology of, for example, D'Inverno\cite{D'Inverno}.} (\ref{9b}) is invariant under spacetime rotations (including boosts) and spacetime inversions (such as reflections) and combinations of these, which make up a group $I_A$ isomorphic to $O(1,3)$. $J_A$ can be partitioned into cosets of the form $l_0 I_A$, so we can always write
\begin{equation} \label{j0-decomp}
j_0 |_A = l_0 |_A i_0 |_A
\end{equation}
where $i_0 \in I_A$.
If we then define
\begin{equation}
\hat{\mathbf{k}}_\kappa |_A = (i_0)_\kappa{}^\xi |_A \; \hat{\mathbf{n}}_\xi |_A \, ,
\end{equation}
we find that
\begin{equation}
(\hat{\mathbf{k}}_\kappa, \hat{\mathbf{k}}_\lambda)_A = \eta_{\kappa \lambda} \, ,
\end{equation}
\begin{equation}
\mathbf{e}_\mu |_A = (l_0)_\mu{}^\kappa |_A \; \hat{\mathbf{k}}_\kappa
\end{equation}
and
\begin{equation} \label{gfromlambda}
g_{\mu \nu} |_A = (l_0)_\mu{}^\kappa |_A \; (l_0)_\nu{}^\lambda |_A \; \eta_{\kappa \lambda} \, .
\end{equation}

\section{Connections and covariant derivatives along a curve \label{curve}}

Having examined the tangent space at a given point $A$, we now want to look at comparing the tangent spaces at different points. To do this, we need to use a connection. 

General relativity uses a particular connection, the Levi-Civita connection, or Christoffel symbol. This has the advantages of being symmetric and being uniquely defined - on a given manifold in a given coordinate system, its components are single-valued at each point. However, when considering frame bases as we are here, it makes more sense to introduce the concepts by starting with connections on a curve, which can be generalized either to the Levi-Civita connection and its associated spin connection, or to those of teleparallelism.

Consider a curve $c(s)$ through $\Omega$ parametrized by the single variable $s$. This may or may not be a geodesic. We take $s$ to be invariant under changes of coordinate. Pick two points on it $A$ and $B$. We define any map between the tangent spaces $T_A \mathcal{M}$ and $T_B \mathcal{M}$ which preserves linearity and the inner product as a `parallel map'. There are an infinite number of these. 

Now choose frame bases at both points, $\hat{\mathbf{n}}_\xi |_A$ and $\hat{\mathbf{n}}_\xi |_B$. Denote the parallel map $\tilde{}$ for which the image of $\hat{\mathbf{n}}_\xi |_A$ is $\hat{\mathbf{n}}_\xi |_B$:
\begin{eqnarray}
\tilde{}: T_A \mathcal{M} &\rightarrow&  T_B \mathcal{M}\\
\tilde{}:\hat{\mathbf{n}}_\xi |_A &\mapsto& \hat{\mathbf{n}}_\xi |_B \label{parallel} \, .
\end{eqnarray}
Then as $\tilde{}$ is a linear map,
\begin{equation}
\tilde{}: \mathbf{e}_\mu |_A \mapsto \tilde{ \mathbf{e}}_\mu = (j_0 |_A \; j_0{}^{-1} |_B)_\mu{}^\nu \mathbf{e}_\nu |_B \, .
\end{equation}

In the teleparallism formalism, this is valid regardless of how close or far apart $A$ and $B$ are. However, we are looking to define a connection. We therefore take $A$ and $B$ to be close to each other (the interval between these events is small). We then note that we can also define parallel maps to and from all the points on $c(s)$ between these points - this set of parallel maps along this section of the curve constitutes a `parallelism'. We choose this such that the transformation $j_0$ from the frame basis to the coordinate basis varies continuously with $s$. (This means that not only must the coordinate basis and the frame basis be related by the same group $J$ all along the curve, but $j_0$ must be in the same connected component of $J$ at all points.) This allows us to carry out a Taylor expansion of $j_0{}^{-1}$ in $s$, giving us

\begin{equation} \label{Taylor}
\tilde{ \mathbf{e}}_\mu = \left(\mathbf{1} + \delta s \left.\left( j_0 \partial_s j_0{}^{-1}\right)_\mu{}^\nu  \right|_A \right) \mathbf{e}_\nu |_B + \mathcal{O}^2 (s) \, ,
\end{equation}
where $\partial_s$ denotes the differential $\frac{\partial}{\partial s}$. 

From the linear nature of the parallel map, we then find the image of any vector $\mathbf{V}$:
\begin{equation} \label{fbpm}
\tilde{}: \mathbf{V} |_A \mapsto \tilde{\mathbf{V}} = V^\nu |_A \mathbf{e}_\nu |_B +  \delta s \; V^\mu |_A \left. \left( j_0 \partial_s j_0{}^{-1} \right)_\mu{}^\nu \right|_A \mathbf{e}_\nu |_B + \mathcal{O}^2 (s) \, .
\end{equation}

The quantity in brackets is our archetypal connection (up to a change in sign):
\begin{equation} \label{G-curve}
\left. \left(\Gamma^{(u)}_s \right)_\mu{}^\nu \right|_A \equiv - \left. \left(j_0 \partial_s j_0^{-1} \right)_\mu{}^\nu \right|_A = \left. \left(\partial_s (j_0) j_0^{-1} \right)_\mu{}^\nu \right|_A \, .
\end{equation}
This is closely related to the Maurer-Cartan form $j_0 \mathrm{d} j_0^{-1}$. We can follow the method of non-linear realizations and use (\ref{j0-decomp}) to decompose it into subgroup and coset space terms:
\begin{equation} \label{WB-lam}
\left. \left(\Gamma^{(u)}_s \right)_\mu{}^\nu \right|_A = - \left. \left( l_0 \, \partial_s l_0{}^{-1} \right)_\nu{}^\mu \right|_A - \left. \left( l_0 \, \omega_s l_0{}^{-1} \right)_\nu{}^\mu \right|_A
\end{equation}
where $\omega_s$ is defined by
\begin{equation}
(\omega_s)_\nu{}^\mu (u) = - \left( i_0 \, \partial_s i_0{}^{-1} \right)_\nu{}^\mu
\end{equation}
This takes values in the Lie algebra of the Lorentz group and is known as the Lorentz connection or spin connection associated with $\Gamma$. We will see in Section (\ref{inertial}) how this is helpful in understanding inertial forces.

Under a change of curvilinear coordinates, from $u^\kappa$ to $u'^\kappa$, we simply replace $j_0$ in these expressions by $j j_0$, where
\begin{equation}
j_\mu{}^\nu = \frac{\partial u^\nu}{\partial u'^\mu} \, ,
\end{equation}
giving us
\begin{equation}
j: \left. \left( \Gamma^{(u)}_s \right)_\mu{}^\nu \right|_A \mapsto \left. \left( \Gamma^{(u')}_s \right)_\mu{}^\nu \right|_A = \left. \left( j \Gamma_s j^{-1} \right)_\mu{}^\nu \right|_A - \left. \left( j \partial_s j^{-1} \right)_\mu{}^\nu \right|_A \, .
\end{equation}

One possible change of coordinates is to the set $x^\xi$ mentioned above, with pseudo-orthonormal basis at $A$. Then $j = j_0^{-1}$, so
\begin{equation} \label{geo-zero}
\left. \left( \Gamma^{(x)}_s \right)_\mu{}^\nu \right|_A = 0 \, .
\end{equation}
If $c(s)$ is a geodesic, then $x$ can have pseudo-orthonormal basis, and $\Gamma^{(x)}_s = 0$, along the entire curve. This, as we shall see when we look at the geodesic equation, is the world line for a freely falling observer in the absence of non-gravitational forces.

More generally, the action of $j$ on a coset $l_0 I$ is
\begin{equation}
j: l_0 I \mapsto l' I = j l_0 I
\end{equation}
(see (\ref{coset-map})), so that
\begin{equation} \label{l'def}
j l_0 = l' i
\end{equation}
and hence
\begin{equation} \label{j-act}
j l_0 i_0 = l' i i_0 \, .
\end{equation}
This means that the connection (\ref{WB-lam}) is transformed to
\begin{equation}
\left. \left(\Gamma'^{(u)}_s \right)_\mu{}^\nu \right|_A = - \left. \left( l' \, \partial_s l'{}^{-1} \right)_\nu{}^\mu \right|_A - \left. \left( l' \, \omega'_s l'{}^{-1} \right)_\nu{}^\mu \right|_A
\end{equation}
where
\begin{equation}
(\omega'_s)_\nu{}^\mu (u) = - \left( i' \, \partial_s i'{}^{-1} \right)_\nu{}^\mu
\end{equation}
and
\begin{equation} \label{i'def}
i' = i i_0 \, .
\end{equation}

We can also look at changing parallelism. Consider a new parallelism $\bar{} \;$, which again preserves orthonormality, so that
\begin{equation}
\bar{} \; : \left. \hat{\mathbf{n}}_\xi \right|_A \mapsto \left. \bar{\mathbf{n}}_\xi \right|_B = i_\xi{}^\rho \left. \hat{\mathbf{n}}_\rho \right|_B \, .
\end{equation}
If $i$ is constant along $c(s)$, $\Gamma_s$ is unaffected. But if $i$ varies with $s$ (we take it to be in the same connected component of $I$ at every point),
\begin{equation}
\bar{} \; : \left. \left( \Gamma_s \right)_\mu{}^\nu \right|_A \mapsto \left. \left( \Gamma'_s \right)_\mu{}^\nu \right|_A - \left. \left( j_0 i \partial_s (i^{-1}) j_0^{-1} \right)_\mu{}^\nu \right|_A  \, .
\end{equation}

We can use (\ref{fbpm}) to define a covariant derivative:
\begin{equation} \label{cdcurve}
D_s V^\nu = \partial_s V^\nu + V^\mu \left( \Gamma_s \right)_\mu{}^\nu \, .
\end{equation}
It is easy to show that this transforms covariantly:
\begin{equation}
j : D^{(u)}_s V^\mu_{(u)} \mapsto D^{(u')}_s V^\mu_{(u')} = D^{(u)}_s V^\nu_{(u)} \left(j^{-1} \right)_\nu{}^\mu \, .
\end{equation}
In the $x$ coordinates, this simply becomes
\begin{equation}
D^{(x)}_s V^\mu_{(x)} = \partial_s V^\mu_{(x)} \, .
\end{equation}

\section{Connections and covariant derivatives across the coordinate neighbourhood \label{chart}}

\subsection{The Weitzenb\"{o}ck and Levi-Civita connections}

It is possible to extend the way we defined $\Gamma$ above to the whole of $\Omega$. Rather than just defining a parallelism - a set of parallel maps - along a curve, we define a parallelism across the whole of $\Omega$. Note that we are not defining it for the whole manifold. This cannot be done for most Riemannian and pseudo-Riemannian manifolds (they are not `parallelisable'). For example, a parallelism can be defined on the two-sphere, which fails at least one point but is valid at all other points. However, this is more than adequate for a coordinate neighbourhood, which may cover, for example, one hemisphere, or even part of a hemisphere. For our manifold, we assume that we can choose $\Omega$ to be small enough that a single parallelism can be used for all of it.

This results in $j_0$ becoming a field over $u^\xi$. We can then define a connection field using the same approach as in (\ref{Taylor}), except we now Taylor expand in each of the curvilinear coordinates; this is known as the Weitzenb\"{o}ck connection:
\begin{equation} \label{WBdefn}
\dot{\Gamma}_{\lambda \nu}{}^\mu (u) \equiv - \left( j_0 \partial_\lambda j_0{}^{-1} \right)_\nu{}^\mu \equiv \left( \partial_\lambda (j_0) j_0{}^{-1} \right)_\nu{}^\mu \, .
\end{equation}

This is not the most general connection. Other rules for parallel transporting a vector exist, which do not take this form. More generally,
\begin{equation} \label{pmap}
\tilde{}: \mathbf{V} |_A \mapsto \tilde{\mathbf{V}} = V^\nu |_A \mathbf{e}_\nu |_B -  \delta u^\lambda \; V^\mu |_A \Gamma_{\lambda \mu}{}^\nu \mathbf{e}_\nu |_B + \mathcal{O} (\delta u)^2 \, .
\end{equation}

The transformation of $\Gamma_{\lambda \mu}{}^\nu$ under a local change of basis is similar to the transformation for $\Gamma_s$, except that we now need to act on the index $\lambda$:
\begin{equation} \label{j-on-conn}
j(u): \Gamma^{(u)}_{\lambda \mu}{}^\nu \mapsto \Gamma^{(u')}_{\lambda \mu}{}^\nu =  j_\lambda{}^\kappa \left( j \Gamma^{(u)}_\kappa j^{-1} \right)_\mu{}^\nu - j_\lambda{}^\kappa \left( j \partial_\kappa j^{-1} \right)_\mu{}^\nu
\end{equation}
where $(\Gamma_\lambda)_\mu{}^\nu \equiv \Gamma_{\lambda \mu}{}^\nu$.

Just as for $\Gamma_s$, we can apply a transformation $j_0^{-1}$ to reduce the Weitzenb\"{o}ck connection to zero - except that we can now do it over the whole of $\Omega$. However, on a curved manifold, the frame bases defined by
\begin{equation}
\hat{\mathbf{n}}_\xi = \left(j_0^{-1} \right)_\xi{}^\mu \mathbf{e}_\mu
\end{equation}
at each point \emph{do not represent the basis for any coordinate system}. This must be the case: if these bases are the basis for a coordinate system, the metric in these coordinates is the Minkowski metric. This is only possible if the manifold is flat and the coordinates are Minkowski coordinates. Thus the price we have to pay for adopting a parallelism across a coordinate neighbourhood is giving up our interpretation of $j_0$ as a Jacobian. (The technical definition of the parallelism is that the map $\tilde{}$ associates these bases through (\ref{parallel}) for any two points $A$ and $B$ on $\Omega$, regardless of distance and independent of path between them.)

However, we can still take $j_0$ to be a Jacobian along a specific geodesic, by using so called `Riemann normal coordinates'. If we consider a point particle moving along a geodesic, we can always base a set of coordinates $x^\xi$ on its rest frame. The geodesic is parametrized by $\tau$, the particle's proper time, which is proportional to $x^0$:
\begin{equation}
x^0 = \mathrm{c} \tau \, .
\end{equation}
These coordinates have pseudo-orthonormal basis along the entire geodesic, and indeed the first derivatives of the metric are zero. By comparison with (\ref{geo-zero}), we therefore have
\begin{equation} \label{Weitzrest}
\left. \dot{\Gamma}^{(x)}_{0 \mu}{}^\nu \right|_{c(\tau)} = 0 \, .
\end{equation}
We look more closely at the physical interpretation of Riemann normal coordinates and the issue of adapting a parallelism to a geodesic in Section \ref{inertial}.

For any connection $\Gamma^{(u)}_{\lambda \mu}{}^\nu$, we may define the covariant derivative of a vector, with components
\begin{equation}
D_\lambda V^\nu = \partial_\lambda V^\nu + V^\mu \Gamma_{\lambda \mu}{}^\nu \, .
\end{equation}
The covariant derivative at a point $A$ is an element of $T_A \mathcal{M} \otimes T^*_A \mathcal{M}$. Under a local change of basis, the inhomogeneous term in the transformation of $\Gamma$ is canceled by the inhomogeneous term in the transformation of $\partial_\lambda V^\mu$. Consequently, $D_\lambda V^\nu$ transforms covariantly:
\begin{equation}
j : D^{(u)}_\lambda V^\mu_{(u)} \mapsto D^{(u')}_\lambda V^\mu_{(u')} = j_\lambda{}^\kappa D^{(u)}_\kappa V^\nu_{(u)} \left(j^{-1} \right)_\nu{}^\mu \, .
\end{equation}
This can be extended in the normal way to tensors of other ranks. 

It is easy to show that any connection for which (\ref{pmap}) preserves the inner product of vectors is metric compatible, that is
\begin{equation}
D_\lambda g^{\mu \nu} = 0 \, .
\end{equation}
However, it is not necessarily symmetric. For example, the Weitzenb\"{o}ck connection is metric compatible, but has a torsion:
\begin{equation}
\dot{T}_{\lambda \mu}{}^\nu = \dot{\Gamma}_{\lambda \mu}{}^\nu - \dot{\Gamma}_{\mu \lambda}{}^\nu \neq 0 \, .
\end{equation}
The only symmetric, metric-compatible connection is the Levi-Civita connection:
\begin{equation}
\mathring{\Gamma}_{\lambda \mu}{}^\nu = \mathring{\Gamma}_{\mu \lambda}{}^\nu = \frac{1}{2} g^{\nu \kappa} \left(\partial_\kappa g_{\lambda \mu} - \partial_\lambda g_{\kappa \mu} - \partial_\mu g_{\kappa \lambda} \right) \, .
\end{equation}
As shown by Pereira\cite{Pereira} and others, any non-symmetric, metric-compatible connection, including the Weitzenb\"{o}ck connection, can be written as the sum of its contorsion and the Levi-Civita connection:
\begin{equation} \label{GGK}
\Gamma_{\lambda \mu}{}^\nu = \mathring{\Gamma}_{\lambda \mu}{}^\nu + K_{\lambda \mu}{}^\nu
\end{equation}
where
\begin{equation}
K_{\lambda \mu}{}^\nu = \frac{1}{2} \left(T_\mu{}^\nu{}_\lambda + T_\lambda{}^\nu{}_\mu - T_{\lambda \mu}{}^\nu \right) \, .
\end{equation}

Now for any geodesic $c(s)$, in the Riemann normal coordinates $x^\xi$, the derivatives of the metric are zero, so 
\begin{equation}
\left. \mathring{\Gamma}^{(x)}_{\lambda \mu}{}^\nu \right|_{c(\tau)} = 0 \, .
\end{equation}
However, away from the geodesic the Levi-Civita connection is non-zero on a curved manifold, even in this coordinate system. Note that incorporating (\ref{Weitzrest}), we have
\begin{equation}
\left. \dot{\Gamma}^{(x)}_{0 \mu}{}^\nu \right|_{c(\tau)} = 
\left. \mathring{\Gamma}^{(x)}_{0 \mu}{}^\nu \right|_{c(\tau)} = 0 \, .
\end{equation}
Thus we see that the timelike part of the Weitzenb\"{o}ck contorsion along a geodesic is zero in Riemann normal coordinates. We look at the physical interpretation of this in Section \ref{inertial}.

Note that the decomposition (\ref{GGK}) is consistent with what we know about flat regions of spacetime. In this case, $\hat{\mathbf{n}}_\xi$ is the coordinate basis for the Minkowski coordinates $x^\xi$. This means that $j_0$ would be a Jacobian matrix across such a region. Consequently, the Weitzenb\"{o}ck connection could be written
\begin{equation}
\dot{\Gamma}_{\lambda \nu}{}^\mu (u) = \partial_\lambda \left(\frac{\partial x^\xi}{\partial u^\nu} \right) (j_0{}^{-1})_\xi{}^\mu \, .
\end{equation}
This has no torsion, so from its definition, the Weitzenb\"{o}ck contorsion tensor would vanish. This, according to (\ref{GGK}), would leave the Weitzenb\"{o}ck connection equal to the Levi-Civita connection. Finally, this would mean that the Levi-Civita connection is pure gauge and could be eliminated by a change of coordinates (to the $x$ coordinates).

\subsection{Field strengths}

The Weitzenb\"{o}ck connection has zero field strength\cite{Pereira,BST}:
\begin{equation}
\partial_\lambda \dot{\Gamma}_{\nu \kappa}{}^\mu - \partial_\nu \dot{\Gamma}_{\lambda \kappa}{}^\mu + \dot{\Gamma}_{\nu \kappa}{}^\rho  \dot{\Gamma}_{\lambda \rho}{}^\mu - \dot{\Gamma}_{\lambda \kappa}{}^\rho  \dot{\Gamma}_{\nu \rho}{}^\mu = 0
\end{equation}
and, as noted above, it can be reduced to zero across $\Omega$ by a local change of basis. The scalar curvature (the Ricci scalar) may be constructed from its torsion tensor\cite{Pereira,Maluf}. For a given coordinate system on a given manifold, this connection is not unique - its definition depends on the parallelism chosen. 

The field strength of the Levi-Civita connection is the Riemann curvature tensor:
\begin{equation}
\mathring{R}^\mu {}_{\kappa \lambda \nu} = \partial_\lambda \mathring{\Gamma}_{\nu \kappa}{}^\mu - \partial_\nu \mathring{\Gamma}_{\lambda \kappa}{}^\mu + \mathring{\Gamma}_{\nu \kappa}{}^\rho \mathring{\Gamma}_{\lambda \rho}{}^\mu - \mathring{\Gamma}_{\lambda \kappa}{}^\rho \mathring{\Gamma}_{\nu \rho}{}^\mu
\end{equation}
and the connection cannot be reduced to zero across $\Omega$ by a local change of basis, except on a flat spacetime. For a given coordinate system on a given manifold, it is unique. The Riemann tensor can also be viewed in terms of the action of the covariant derivatives on a vector field:
\begin{equation}
\left[ \mathring{D}_\kappa, \mathring{D}_\rho \right] V^\xi = \mathring{R}^\xi{}_{\lambda \kappa \rho} V^\lambda \, .
\end{equation}

\subsection{Lorentz connections}

Each metric-compatible connection $\Gamma_{\lambda \nu}{}^\mu$ has an associated Lorentz connection or spin connection, taking values in the Lie algebra of the Lorentz group\cite{Pereira}. This means that at least two of its indices must be frame indices. It therefore has two forms, one of which has all three indices as frame indices, while the other has two frame indices and one coordinate index. In the formalism of this paper, the Lorentz connection with three frame indices is considered to be the usual connection in the frame basis. The frame basis at a point $A$ is the basis at that point for some set of Riemann normal coordinates $x^\xi$, so we can write this connection at this point as $\left. \Gamma^{(x)}_{\lambda \kappa}{}^\mu \right|_A$. The form with two frame indices and one coordinate index is considered to be in a mix of two different bases. We shall write this as follows:
\begin{equation}
\left. \omega_\mu{}^{\xi \rho} \right|_A M_{\xi \rho} \in \mathcal{I_A} \, ,
\end{equation}
where the first index is taken to be a coordinate index and the last two are frame indices.

If we choose a frame $\hat{\mathbf{n}}_\mu$ at $A$ related to the coordinate basis by (\ref{j0-action}) where $j_0$ can be decomposed using (\ref{j0-decomp}), any metric-compatible connection in the coordinate basis can be related to a Lorentz connection as follows:
\begin{equation} \label{Lor2gam}
\Gamma^{(u)}_{\mu \lambda}{}^\nu = (l_0)_\lambda{}^\kappa \omega_{\mu \kappa}{}^\rho (l_0^{-1})_\rho{}^\nu + (l_0)_\lambda{}^\kappa \partial_\mu (l_0^{-1})_\kappa{}^\nu \, ,
\end{equation}
where the first index of $\omega_{\mu \kappa}{}^\rho$ is taken to be in the coordinate basis, with frame indices raised and lowered using $\eta^{\xi \kappa}$ and $\eta_{\xi \kappa}$. This equation can be inverted to give:
\begin{equation}
\omega_{\mu \xi}{}^\rho = (l_0^{-1})_\xi{}^\lambda \Gamma^{(u)}_{\mu \lambda}{}^\nu (l_0)_\nu{}^\rho + (l_0^{-1})_\xi{}^\lambda \partial_\mu (l_0)_\lambda{}^\rho \, .
\end{equation}
However, even for a given connection $\Gamma$, this Lorentz connection is not unique: any local change of frame $i(u)$ (including $i_0 (u)$) results in another Lorentz connection. $\omega_{\mu \xi}{}^\rho$ transforms under a local change of frame according to:
\begin{equation} \label{Lgauge}
i(u): \omega_{\mu \xi}{}^\rho \mapsto \omega'_{\mu \xi}{}^\rho =  \left( i \omega_\mu i^{-1} \right)_\xi{}^\rho + \left( i \partial_\mu i^{-1} \right)_\xi{}^\rho \, .
\end{equation}
For the Weitzenb\"{o}ck connection, (\ref{Lor2gam}) amounts to a Cartan decomposition using (\ref{j0-decomp}):
\begin{equation} \label{G-decomp}
\dot{\Gamma}^{(u)}_{\mu \lambda }{}^\nu = (l_0)_\lambda{}^\xi \dot{\omega}_{\mu \xi}{}^\rho (l_0^{-1})_\rho{}^\nu + \dot{\rho}_{\mu \lambda}{}^\nu 
\end{equation}
where
\begin{equation} \label{Weitzspin}
\dot{\omega}_{\mu \xi}{}^\rho = (i_0 \partial_\mu i_0^{-1})_\xi{}^\rho 
= - [(\partial_\mu i_0) i_0^{-1}]_\xi{}^\rho
\end{equation}
and
\begin{equation}
\dot{\rho}_{\mu \xi}{}^\rho = (l_0 \partial_\mu l_0^{-1})_\xi{}^\rho 
= - [(\partial_\mu l_0) l_0^{-1}]_\xi{}^\rho \, .
\end{equation}
(This is the extension of (\ref{WB-lam}) to the coordinate neighbourhood $\Omega$.) (\ref{Weitzspin}) means that the Weitzenb\"{o}ck spin connection can be reduced to zero everywhere by a local change of frame, whereas the Levi-Civita spin connection cannot\cite{Pereira}. The choice of gauge which reduces it to zero is often called the Weitzenb\"{o}ck gauge and $\dot{\rho}_{\mu \xi}{}^\rho$ is generally viewed in the literature as the Weitzenb\"{o}ck connection in this Weitzenb\"{o}ck gauge.

This Cartan decomposition can be applied to the transformation of the Weitzenb\"{o}ck connection under $j$, given by (\ref{j-on-conn}). The results are again similar to those for $\Gamma_s$, but with an extra transformation of the third index:
\begin{equation} \label{jonW}
j: \dot{\Gamma}^{(u)}_{\lambda \mu}{}^\nu \mapsto 
\dot{\Gamma'}^{(u)}_{\kappa \mu}{}^\nu = - j_\kappa{}^\lambda [\left( l' \, \partial_\lambda l'{}^{-1} \right) + \left( l' \, \dot{\omega}'_\lambda l'{}^{-1} \right)]_\nu{}^\mu
\end{equation}
where
\begin{equation} \label{jonWS}
(\dot{\omega}'_\lambda)_\nu{}^\mu (u) \equiv \dot{\omega}'_{\lambda \nu}{}^\mu (u) = - \left( i' \, \partial_\lambda i'{}^{-1} \right)_\nu{}^\mu \, .
\end{equation}
and $l'$ and $i'$ are again given by (\ref{l'def}) and (\ref{i'def}).

\medskip

It will be noted that $\dot{\rho}_{\mu \xi}{}^\rho$ depends only on $l_0$ and its derivatives. Given that the metric only carries the degrees of freedom of $l_0$, it may be wondered whether this term is related to the Levi-Civita connection. Actually, it can be shown that it is. From (\ref{gfromlambda}) (when applied across the chart), we have
\begin{eqnarray}
\partial_\kappa g_{\mu \nu} &=& (\partial_\kappa (l_0)_\mu{}^\xi (l_0)_\nu{}^\rho
+ (l_0)_\mu{}^\xi \partial_\kappa (l_0)_\nu{}^\rho ) \; \eta_{\xi \rho} \\
&=& [- (l_0 \partial_\kappa l_0^{-1} )_\mu{}^\lambda (l_0)_\lambda{}^\xi (l_0)_\nu{}^\rho - (l_0 \partial_\kappa l_0^{-1} )_\nu{}^\lambda (l_0)_\mu{}^\xi (l_0)_\lambda{}^\rho] \; \eta_{\xi \rho} \\
&=& - (\dot{\rho}_\kappa)_{\left\lbrace \mu\right.}{}^\lambda g_{\lambda \left. \nu \right\rbrace} \, ,
\end{eqnarray}
where to get the second line, we have inserted $l_0 l_0^{-1}$ into both terms then used $l_0^{-1} \partial_\kappa l_0 = - \partial_\kappa (l_0^{-1}) l_0$. The braces round the indices in the last line represent symmetrization on those indices. From this we find that the Levi-Civita connection is
\begin{eqnarray}
\mathring{\Gamma}_{\kappa \nu}{}^\rho &=& \frac{1}{2} g^{\rho \mu} (\partial_\kappa g_{\mu \nu} + \partial_\nu g_{\mu \kappa} - \partial_\mu g_{\kappa \nu}) \\
&=& \frac{1}{2} g^{\rho \mu} (- \dot{\rho}_{[\kappa \mu] }{}^\lambda g_{\lambda \nu} 
- \dot{\rho}_{[\kappa \nu] }{}^\lambda g_{\mu \lambda} + \dot{\rho}_{\left\lbrace \mu \nu \right\rbrace}{}^\lambda g_{\kappa \lambda})
\end{eqnarray}
where the square brackets round the indices represent antisymmetrization on those indices. Thus the Levi-Civita connection depends only on $\dot{\rho}_{\mu \xi}{}^\rho$ and the metric - we can thus see why it has become so popular to carry out calculations in the Weitzenb\"{o}ck gauge.

Note that when $\dot{\rho}_{\mu \xi}{}^\rho$ is symmetric on its lower indices, it reduces to the Levi-Civita connection (with one index raised and another one lowered). However, there does not appear to be anything requiring this to be the case. In general, both terms in (\ref{G-decomp}) carry torsion, so this decomposition is a different one to that given in (\ref{GGK}). More precisely, all three terms in (\ref{GGK}) for the Weitzenb\"{o}ck connection are dependent on $\dot{\rho}$, but only the first and last terms carry an $\dot{\omega}$-dependence.

\section{Using the coset decomposition to study Lorentz gauge transformations and  inertial forces on an observer \label{inertial}} 

\subsection{Inertial forces are not induced by Lorentz transformations}

As mentioned in Section \ref{OandS}, there has recently been considerable debate about the role of the Lorentz gauge transformation. As we will show below, the coset decomposition is the appropriate framework in which to address this question.

However, it is first worth correcting an issue which seems to be causing some confusion. It often seems to be assumed that a change of reference frame which induces inertial forces, such as centrifugal forces, can be represented by a local Lorentz transformation. It seems a natural assumption to make, following the logic that a rotation of coordinates is a Lorentz transformation, so a rotation of coordinates that changes over time is a local Lorentz transformation. However, this is not actually the case; furthermore, we do not need the apparatus of teleparallelism to show this. The flaw in the above logic is that time itself is one of the coordinates and this results in a non-pseudo-orthogonal transformation, as we now show.

Consider two coordinate systems, $x^\mu$ and $x'^\mu$. For convenience, we will place them on a flat spacetime and make $x'^\mu$ the Minkowski coordinates. If $x'^\mu$ is rotated with respect to $x^\mu$ through an angle $\theta$ in the $x$-$y$ plane, then
\begin{equation}
x' = x \cos \theta + y \sin \theta, \hspace{8mm} y' = - x \sin \theta + y \cos \theta, \hspace{8mm} z' = z, \hspace{8mm} t' = t \, .
\end{equation}
If we now let $\theta$ change over time, so that $x'^\mu$ rotates with respect to $x^\mu$ at a rate of $k$ radians per second,
\begin{equation}
x' = x \cos kt  + y \sin kt, \hspace{7mm} y' = - x \sin kt + y \cos kt, \hspace{7mm} z' = z, \hspace{7mm} t' = t \, ;
\end{equation}
it is then a simple calculation to show that
\begin{eqnarray}
ds'^2 &=& c^2 dt'^2 - dx'^2 - dy'^2 - dz'^2 \\
&=& (c^2 - k^2 x^2 - k^2 y^2) dt^2 - dx^2 - dy^2 - 2ky \, dx \, dt - 2kx\,  dy \, dt  - dz^2 \, . \nonumber \\
\end{eqnarray}
Given that $ds^2$ is an invariant under changes of coordinates, it is clear that the metric is not preserved (it is no longer the Minkowski metric $\eta_{\mu \nu}$), so the transformation cannot be pseudo-orthogonal.

From the GR perspective, it makes sense that the introduction of an inertial force requires a change of metric. Without this, the Levi-Civita connection remains zero. The experience of an observer following a path $u(\tau)$ is determined by the geodesic equation
\begin{equation} \label{geodesic}
\frac{\partial^2 u^\mu}{\partial \tau^2} + \mathring{\Gamma}_{\kappa \lambda}{}^\mu \frac{\partial u^\kappa}{\partial \tau} \frac{\partial u^\lambda}{\partial \tau} = 0
\end{equation}
and a zero Levi-Civita connection means that the observer feels no force.

\subsection{Choices of parallelism and coordinates, and the forces experienced by an observer}

Let us consider two situations, familiar to students of general relativity. 

\smallskip

Situation 1 is that of a free observer situated in a flat spacetime (this could, for example, represent deep space). A natural choice of coordinates for this spacetime would be Minkowski coordinates, which can be used over the entire spacetime (at least, for the extent in each direction in which it remains flat). The coordinate basis is then also a frame basis at every point.

By associating this basis at a point $A$ with the basis at every other point, (so that they are images of each other under parallel transport), we can define a parallelism. We could say that this parallelism is `adapted to' the Minkowski coordinates on this spacetime. With these choices, $j_0$ reduces to the identity: $\mathbf{e}_\mu$ coincides with $\hat{\mathbf{n}}_\mu$.

However, other choices are available. While retaining the Minkowski coordinates, we could change to a different parallelism, which is not adapted to these coordinates, using (\ref{Lgauge}). This may seem a perverse thing to do, but the paralellism is just a choice. $l_0$ remains the identity, so $\mathbf{e}_\nu = \hat{\mathbf{k}}_\nu = i_\nu{}^\mu \hat{\mathbf{n'}}_\mu$.

Alternatively, we could change to a set of curvilinear coordinates. This changes the coordinate basis according to (\ref{j0-action}).

\smallskip

Situation 2 is that of an observer following a geodesic on a curved spacetime - that is, falling freely in a gravitational field. A natural choice of coordinates for this situation are those of the observer's local inertial frame - these are Riemann normal coordinates whose basis is orthonormal along the geodesic.

We can choose a parallelism which is adapted to these coordinates on this geodesic, so that again $j_0$ is the identity element along the geodesic. \emph{However, on a curved spacetime, there is no parallelism which is adapted to all the geodesics across an extended multi-dimensional region of the spacetime.} Thus $j_0$ cannot be the identity everywhere. (Indeed, $l_0$ cannot be the identity everywhere, as this would make the metric the Minkowski one everywhere.)

Again, other choices are available. While retaining the Riemann normal coordinates, we could choose a different parallelism, for example one adapted to a world line which crosses that of our observer.

Alternatively, we could change to a different set of curvilinear coordinates - for example, the rest frame for an observer whose path crosses that of our observer, or the rest frame of the mass distribution causing the curvature.

\smallskip

The question is then what impact these changes have on the experience of the observer. In GR, this is determined by the geodesic equation (\ref{geodesic}). The crucial point is that this only depends upon the Levi-Civita connection; it is independent of the choice of parallelism. (We could use (\ref{GGK}) to rewrite the connection term above, but then it would depend upon both the Weitzenb\"{o}ck connection \emph{and} its contorsion.) This must also be the case for any theory in which the geodesic equation holds - including TEGR, as it is equivalent to GR at the level of physical observations.

This can be understood physically as follows. In general relativity, and hence TEGR, the observer `lives' in the frame of reference defined by the coordinates $u^\mu$. This is independent of the choice of frame bases along their world line - that is, the parallelism used - which is simply a choice for the researcher carrying out calculations. This should also hold for modified theories. The parallelism is a choice of Lorentz gauge chosen by the analyst to make calculations easier. The experience of the observer should not change if this gauge choice changes, as long as the coordinate basis $\mathbf{e}_\mu$ remains the same\footnote{It should be stressed that this is for a classical, pointlike observer, as in the Einstein lift experiments - if the analysis were extended to incorporate field configurations which take values in a Hilbert space, then it could be that measurements of quantum numbers could distinguish between parallelisms. Such analysis lies outside the scope of this paper.}.

A change of coordinates, however, represents a change of reference frame - from one observer to another - and this changes the Levi-Civita connection. Hence one observer will not necessarily feel the same force as another. 

In situation 1, in Minkowski coordinates, the Levi-Civita connection vanishes. The observer therefore feels no force. A change of coordinate system from Minkowski coordinates to a curvilinear system induces a non-zero Levi-Civita connection, which is felt as inertial forces. 

In situation 2, in Riemann normal coordinates, the Levi-Civita connection vanishes along the observer's world line, so the observer feels no force. However, it is non-zero away from this, representing the tidal effects of gravity. On changing coordinates, the Levi-Civita connection becomes non-zero. This could, for example, be changing to the reference frame of an observer on the surface of a solid gravitating body, at rest with respect to that body. In this case, the Levi-Civita connection would represent the gravitational force as felt by the observer. Alternatively, we could choose a more exotic reference frame, such as one representing someone spinning with respect to the freely-falling observer. The Levi-Civita connection along the world line would then represent the centrifugal force felt by the new observer; beyond their world line it would combine this force with the gravitational tidal effects.

The Levi-Civita connection, therefore, recognizes both gravity and inertial forces and does not distinguish between them at the local level, as observed by Aldrovandi \emph{et al}\cite{ALP}. One has to consider how it varies over spacetime to distinguish between them. This is done through the Riemann tensor, which is zero when spacetime is flat and nonzero when it is curved. This means that when gravity is reduced to zero, the Levi-Civita connection becomes pure gauge - it is zero in Minkowski coordinates, and in any other coordinate system it reduces to the form (\ref{WBdefn}), where $j_0$ is the Jacobian matrix from the Minkowski coordinates to the curvilinear ones. As the deviation between the Weitzenb\"{o}ck and Levi-Civita connections is the Weitzenb\"{o}ck contorsion tensor, this must reduce to zero for Riemann-flat spacetime (in any coordinates). Conversely, as the field strength of the Weitzenb\"{o}ck connection is always zero, a spacetime with Riemann curvature will always have non-zero Weitzenb\"{o}ck contorsion tensor. Thus the Weitzenb\"{o}ck contorsion tensor also has the property that it can be used to determine whether a gravitational field is present.

Researchers have found that in some situations, it is easier to work with the Weitzenb\"{o}ck torsion or contorsion tensors than the Riemann tensor, as these depend only on the first derivatives of the `tetrad fields' - that is, of the matrices relating different bases. The price of this is that it becomes difficult to maintain general covariance unless one keeps track of not just the $l$ degrees of freedom but also the $i$ degrees of freedom - despite the fact that these extra degrees of freedom do not affect physical observables.

\subsection{How changes of coordinate and parallelism both induce Lorentz gauge transformations}

Having looked at the impact of these changes on the Levi-Civita connection, we can also look at how they affect the two terms in the Cartan decomposition of the Weitzenb\"{o}ck connection. A change of parallelism is given by a local Lorentz transformation of the frame basis, $i(u)$. This means that the coordinate basis $\mathbf{e}_\mu$ is related to the new frame basis $\hat{\mathbf{n'}}_\xi$ by
\begin{equation}
\mathbf{e}_\mu = (l_0 i_0 i)_\mu{}^\xi \hat{\mathbf{n'}}_\xi \, .
\end{equation}
This changes the Lorentz connection according to (\ref{Lgauge}) but has no impact on $\dot{\rho}_{\mu \nu}{}^{\xi}$.

A change of coordinates, on the other hand, in general\footnote{There will be some specific changes of coordinates which do not affect either or both of these terms. If the action of a Jacobian matrix $j$ leaves $l_0$ invariant, it must leave both the components of the metric and $\dot{\rho}_{\mu \nu}{}^{\xi}$ invariant. This is not exactly the same as an isometry, which leaves the \emph{functional form} of the metric invariant - this distinction is explained in Section \ref{point}. Transformations which leave the functional forms of these quantities invariant have been studied by Hohmann \emph{et al}\cite{HJKP}, albeit in the tetrad formulation, without the insights contained in this section. The relation between these two types of invariance is unclear, at least within the context of the coset formulation presented here, and is likely to be worthy of further examination.} affects both of these terms, as shown in (\ref{jonW}) and (\ref{jonWS}). However, it is only the change of $\dot{\rho}_{\mu \nu}{}^{\xi}$ which is felt by the observer, as this is the only term which contributes to the Levi-Civita connection. The Lorentz connection remains a gauge choice that has no physical impact on the observer. However, this gauge choice does change with the coordinates. This means, in particular, that the Weitzenb\"{o}ck gauge is specific to the choice of coordinates - one may choose to carry out calculations in the Weitzenb\"{o}ck gauge, but when the coordinates are changed, the system will no longer be in this gauge.

Naturally, the preceding paragraph includes coordinate transformations for which the Jacobian matrix is pseudo-orthogonal. In this case, we can replace $j$ with $i$. Then in place of (\ref{j-act}) we have
\begin{equation}
i l_0 i_0 = i l_0 i^{-1} i i_0 = l' i i_0
\end{equation}
so that the $l'$ in (\ref{jonW}) is given by
\begin{equation}
l' = i l_0 i^{-1} \, ,
\end{equation}
while the $i$ in (\ref{i'def}), which contributes to the $i'$ in (\ref{jonWS}), is precisely the Jacobian matrix for the change of curvilinear basis.

Finally, we can consider the case where
\begin{equation}
\left. l_0  \right|_{c(\tau)} = 0
\end{equation}
- that is, the metric is the Minkowski one along the observer's world line. This means that the observer is in free fall. (Note that we are not restricting $l_0$ away from the world line - it may be non-zero away from it, which means that we may have gravitational forces present.) We do not have to have a parallelism adapted to this world line, so we can still have
\begin{equation}
\left. i_0  \right|_{c(\tau)} \neq 0 \, .
\end{equation}
This means that the Lorentz connection can be non-zero, but $\dot{\rho}_{\mu \xi}{}^\nu$ must be zero. In this case, the Jacobian matrix for an inertial coordinate transformation acts directly on $i_0$, with the action (\ref{i'def}). It therefore changes the Lorentz connection, but the metric remains the Minkowski one along the world line.

\section{Translations \label{Transl}}

\subsection{Translations as coordinate transformations}

As mentioned in Section \ref{Intro}, there has recently been a vigorous debate about whether teleparallel gravity constitutes a gauge theory of translations. Here, we see how the methodology of this paper allows us to study this issue. Our starting point is the formula for coordinate transformations, (\ref{ct}). Consider the subset for which all coefficients after the first two terms are zero - that is, the (global) inhomogeneous linear transformations:
\begin{equation}
u'^\xi = f^\xi + f^\xi{}_\rho u^\rho \, .
\end{equation}
For these, the Jacobian matrix is
\begin{equation}
\frac{\partial u'^\xi}{\partial u^\kappa} = f^\xi{}_\kappa \, .
\end{equation}
From this, we can immediately see that such a transformation preserves the Minkowski metric if and only if $f^\xi{}_\kappa$ is pseudo-orthogonal. Such transformations, with the general form
\begin{equation}
u'^\xi = f^\xi + i^\xi{}_\rho u^\rho
\end{equation}
comprise the Poincar\'{e} group. Amongst these are the global translations
\begin{equation} \label{translate}
u'^\xi = f^\xi + u^\xi
\end{equation}
for which the Jacobian matrix is a Kronecker delta. This means that under the homomorphism (\ref{Jac-an}) from the covariance group into $J_A$, the translations lie in the kernel, so bases and vector components are untransformed. (Note that for vector \emph{fields}, here we are talking about the values of the vector components at a given point, not their functional forms - their expressions in terms of the coordinates, which will obviously change.)

We now want to consider what happens when the translation parameters are made spacetime-dependent. First, we consider the transformation of the basis on the tangent space. This transforms by contraction with the inverse Jacobian matrix, which is
\begin{equation}
\left. \frac{\partial u^\xi}{\partial u'^\kappa} \right|_A = \delta^\xi_\kappa - \left. \frac{\partial f^\xi}{\partial u'^\kappa} \right|_A \, ,
\end{equation}
so
\begin{equation}
f(u^\rho) : \mathbf{e}_\kappa |_A \mapsto \mathbf{e}'_\kappa |_A = \mathbf{e}_\kappa |_A - \left. \frac{\partial f^\xi}{\partial u'^\kappa} \right|_A \mathbf{e}_\xi |_A \, .
\end{equation}
The partial derivative operator transforms in the same way (indeed, it can be seen as a representation of the basis):
\begin{equation} \label{operator}
f (u^\rho) : \partial_\kappa |_A \mapsto \partial'_\kappa |_A = \partial_\kappa |_A - (\partial_\kappa f^\xi) |_A \, \partial_\xi |_A \, .
\end{equation}
Thus for a scalar field $\phi$,
\begin{equation}
f (u^\rho) : \partial_\kappa \phi |_A \mapsto \partial'_\kappa \phi |_A = \partial_\kappa \phi |_A - (\partial_\kappa f^\xi) |_A \, \partial_\xi \phi |_A
\end{equation}
- that is, we have a minimal coupling to the 16 variables of $\partial_\kappa f^\xi |_A$.

What we are doing here is to view the action of the general linear group from a new perspective. We previously considered the action of $j$ by contraction on the frame basis, (\ref{basis-change}). We now consider the displacement of the basis under this action:
\begin{equation}
\delta \mathbf{e}_\kappa |_A = \mathbf{e}'_\kappa |_A - \mathbf{e}_\kappa |_A 
= (j_\kappa{}^\xi |_A - \delta_\kappa{}^\xi) \, \mathbf{e}_\xi |_A 
\end{equation}
or in terms of the translation parameters $f^\xi$:
\begin{equation}
\delta \mathbf{e}_\kappa |_A = - \partial_\kappa f^\xi |_A \mathbf{e}_\xi |_A \, .
\end{equation}
Thus we see that the local translations of the coordinates induce \emph{the same transformations of the tangent space} as described by the action of the general linear group. They are related by
\begin{equation} \label{j-f}
j_\kappa{}^\xi |_A - \delta_\kappa{}^\xi = - \partial_\kappa f^\xi |_A
\end{equation}
and therefore contain the same 16 degrees of freedom. As mentioned in Section \ref{Intro}, this differs from the conclusion of Hohmann \emph{et al}\cite{KHZ} that translations are generated by an \emph{Abelian subgroup} of the general linear group. Instead, we find that \emph{all} general linear transformations \emph{of the basis at $A$} may be expressed in terms of the local values of derivatives of translation parameters. This implies that some of the changes of bases induced by local translations correspond to local Lorentz transformations. By combining (\ref{j-f}) with the definition of a pseudo-orthogonal transformation in $T_A \mathcal{M}$,
\begin{equation}
j_\xi{}^\kappa |_A j_\rho{}^\lambda |_A \eta_{\kappa \lambda} = \eta_{\xi \rho} \, ,
\end{equation}
it emerges that $f^\xi$ represents such a transformation if and only if
\begin{equation}
\eta_{\kappa \lambda} (\partial_\xi f^\kappa) |_A (\partial_\rho f^\lambda) |_A = \eta_{\kappa \rho} (\partial_\xi f^\kappa) |_A + \eta_{\xi \lambda} (\partial_\rho f^\lambda) |_A \, .
\end{equation}
(Note that the very meaning of translations on a curved space is rather different from the simple understanding of them on a flat space. $f^\xi$ represents a quantity which is added to a chosen curvilinear coordinate - this could be, for example, an angular variable.) 

While any change of basis $j$ \emph{in a given tangent space} may be represented in terms of the translation parameters $f^\xi$ according to (\ref{j-f}), it is not true that any change of basis \emph{across a coordinate neighbourhood}, $j(u)$, may be represented in this way. If a field-valued translation induces a change of basis $j_t (u)$ across a coordinate neighbourhood, its non-trivial part will take the form of a derivative:
\begin{equation} \label{j_t}
(j_t)_\kappa{}^\xi - \delta_\kappa{}^\xi = - \partial_\kappa f^\xi
\end{equation}
This puts a constraint on its form as a function of the coordinates $u^\mu$. (Therefore, rather than the subgroup structure described by Hohmann \emph{et al}, the functions $j_t (u)$ are actually a subset of all of the functions $j(u)$.)  In particular, we see that
\begin{equation}
\partial_\mu (j_t)_\kappa{}^\xi = - \partial_\mu \partial_\kappa f^\xi \, .
\end{equation}
This has a significant implication when $j_t$ is the Jacobian matrix relating the coordinate basis to the frame basis, $j_0$, as this then becomes
\begin{equation}
\partial_\mu (j_0)_\kappa{}^\xi = - \partial_\mu \partial_\kappa (f_0)^\xi \, .
\end{equation}
As this second derivative is symmetric on its lower indices, the Weitzenb\"{o}ck connection (\ref{WBdefn}) has no torsion\cite{D'AGHZ}. As we saw in the last section, this can only happen in flat spacetime.

Our conclusion regarding $j_0$ is therefore (in general accord with Hohmann \emph{et al}): \emph{the Weitzenb\"{o}ck connection can only be constructed from translation parameters in a flat region of spacetime}. This can be understood also in terms of the following argument. $j_0$ relates the curvilinear coordinate basis $\mathbf{e}_\kappa$ to the frame basis $\hat{\mathbf{n}}_\kappa$, while $f^\xi$ relates two coordinate systems. If $j_0$ can be expressed in terms of such translation parameters $(f_0)^\xi$, $\hat{\mathbf{n}}_\kappa$ must be the basis corresponding to a coordinate system. This can only happen for an extended region of spacetime (that is, beyond just a geodesic) if that region is flat. In such a case, we may write
\begin{equation} \label{j0-f}
(j_0)_\kappa{}^\xi = \delta_\kappa{}^\xi - \partial_\kappa (f_0)^\xi
\end{equation}
where
\begin{equation}
(f_0)^\xi = u^\xi - x^\xi \, .
\end{equation}

\medskip

Having looked at the action of local translations on the basis, we now want to turn to the action on the components of a vector field. Under a coordinate transformation, $\mathbf{V} |_A \in T_A \mathcal{M}$ remains $\mathbf{V} |_A \in T_A \mathcal{M}$; however, as we have seen above, it can be decomposed in different bases on that tangent space. Let us say that in two such bases, it has components $V^\mu |_A$ and $V'^\mu |_A = V^\mu |_A + \delta V^\mu |_A$. As $\mathbf{e}_\mu |_A$ is a complete linear basis for the tangent space, any other basis vectors can be related to it by (\ref{basis-change}). Thus
\begin{equation}
(V^\mu + \delta V^\mu) |_A \, \mathbf{e}'_\mu |_A = V^\lambda |_A (j^{-1} )_\lambda{}^\mu |_A \mathbf{e}'_\mu |_A \, ,
\end{equation}
from which we find
\begin{equation}
\delta V^\mu |_A  = V^\lambda |_A [(j^{-1})_\lambda{}^\mu |_A - \delta_\lambda^\mu] \, .
\end{equation}
Thus any displacements of the vector components induced by coordinate transformations have the original vector components as a factor.  

Note that one cannot induce inhomogeneous transformations - displacements of the vector components - from a coordinate transformation. These just change the basis under which the value of a vector at a given point is decomposed into components. However, translations can be applied as \emph{point transformations}\footnote{We are using the terminology of D'Inverno\cite{D'Inverno} here.} to any field to generate such transformations. We comment very briefly on this in the next subsection. 

We now need to find an expression for $j_t{}^{-1}$ in terms of the translation parameters. We do this using the local version of (\ref{translate}):
\begin{equation}
(j_t{}^{-1})_\kappa{}^\xi = \frac{\partial u'^\xi}{\partial u^\kappa} = \delta^\xi_\kappa + \frac{\partial f^\xi}{\partial u^\kappa} \, .
\end{equation}
Similarly, for a flat region of spacetime or a geodesic in a curved spacetime,
\begin{equation}
(j_0{}^{-1})_\kappa{}^\xi = \frac{\partial u^\xi}{\partial x^\kappa} = \delta_\xi{}^\kappa + \frac{\partial f_0^\xi}{\partial x^\kappa}
\end{equation}
where $x^\kappa$ are the Minkowski or Riemann normal coordinates.
Note that the non-trivial term on the right-hand side here is not that we have in (\ref{operator}), as the derivative is with respect to $x^\kappa$. We can sort this out with an iterative procedure:
\begin{eqnarray}
(j_0{}^{-1})_\kappa{}^\xi = \frac{\partial u^\xi}{\partial x^\kappa} &=& \delta^\xi_\kappa + \frac{\partial f_0^\xi}{\partial u^\lambda} \frac{\partial u^\lambda}{\partial x^\kappa} \\
&=& \delta^\xi_\kappa + \frac{\partial f_0^\xi}{\partial u^\lambda} \left( \delta^\lambda_\kappa + \frac{\partial f_0^\lambda}{\partial u^\mu} \frac{\partial u^\mu}{\partial x^\kappa} \right) \\
&=& \delta^\xi_\kappa + \partial_\kappa f_0^\xi + (\partial_\kappa f_0^\lambda) (\partial_\lambda f_0^\xi) + \ldots  \label{j0-1exp}
\end{eqnarray}
and similarly
\begin{equation} \label{j-1exp}
((j_t){}^{-1})_\xi{}^\kappa = \delta^\xi_\kappa + \partial_\kappa f^\xi + (\partial_\kappa f^\lambda) (\partial_\lambda f^\xi) + \ldots \, .
\end{equation}
The implications of the convergence criteria for these series remains an open question\footnote{It has come to the author's attention at a late stage in the preparation of this paper that this perturbative expansion has also been carried out by Beltran Jiminez and Dialektopoulos\cite{BJD}.}.

Equation (\ref{j_t}) and (\ref{j-1exp}) can then be substituted into any equations in the earlier sections which contain $j$ and/or $j^{-1}$, to get equations containing the translation parameters. Similarly, (\ref{j0-f}) and (\ref{j0-1exp}) can be substituted into any equations for flat spacetime or a geodesic (\emph{but only for these}) which contain $j_0$ and/or $j_0^{-1}$. For example, the covariant derivative (\ref{cdcurve}) along a geodesic is
\begin{equation}
D_\tau V^\nu = \partial_\tau V^\mu + V^\mu \left( \Gamma_\tau \right)_\mu{}^\nu \, ,
\end{equation}
where $\tau$ denotes proper time and 
\begin{equation}
\left(\Gamma^{(u)}_\tau \right)_\mu{}^\nu = \left(\partial_\tau (j_0) j_0^{-1} \right)_\mu{}^\nu = - \partial_\tau \partial_\mu f_0^\xi \left(\delta^\nu_\xi + \partial_\xi f_0^\nu + (\partial_\xi f_0^\rho) (\partial_\rho f_0^\nu) + \ldots \right) \, ,
\end{equation}
while a covariant derivative with respect to the coordinates (involving any connection) transforms according to
\begin{eqnarray}
f^\xi : D^{(u)}_\lambda V^\mu_{(u)} \mapsto D^{(u')}_\lambda V^\mu_{(u')} = (\delta_\lambda{}^\kappa - \partial_\lambda f^\kappa) D^{(u)}_\kappa V^\nu_{(u)} \left(\delta^\mu_\nu + \partial_\nu f^\mu \right. \nonumber \\
\left. + (\partial_\nu f^\rho) (\partial_\rho f^\mu) + \ldots \right) \, .
\end{eqnarray}

\subsection{Translations as point transformations \label{point}}

In everything preceding this subsection, we have focused on coordinate transformations. 

Each component of a vector field is itself field-valued and its functional form may be very different in two coordinate systems. For example, if the $x$-component of a vector field is
\begin{equation}
V^1 = x^2 + 3y \, ,
\end{equation}
then at a point $A$ with coordinates $(x=2,y=3)$ it has a value of 13. Under the change of coordinates 
\begin{equation}
(x,y) \mapsto (x',y') = (x+3,y) \, ,
\end{equation}
it takes the form
\begin{equation}
V^1 = x'^2 - 6x' + 9 + 3y'
\end{equation}
- a very different functional form, but note that the basis has not changed. The point $A$ now has coordinates $(x'=5,y'=3)$, but $V^1$ still has the value 13 at this point. This naturally goes for scalar, spinor and tensor fields as well. 

A `point translation' is either an active transformation - a physical displacement of the field across spacetime - or, equivalently, the change to its functional form which results from applying the inverse transformation to the underlying coordinates. (In the above case, either shifting the field configuration three units to the left, or shifting the axes three units to the right.) The resulting transformation of the field is found through a Taylor expansion:
\begin{equation}
V^1 (x') = V^1 (x) + \delta x \, \partial_x V + \mathcal{O}^2 (\delta x) \, ,
\end{equation}
which clearly allows an inhomogeneous transformation of the vector component. The generator of such a transformation is the partial derivative operator, which is proportional to the quantum mechanical momentum operator. These transformations seem to be the ones used by B\"{o}hmer \emph{et al}\cite{BCKPVDH}, who use a Noether-type argument that teleparallel gravity may be described as a gauge theory of translations. (It is not clear that these are the transformations used by Pereira and Obukhov\cite{PO}, who seem to postulate a translation action on a vector at the outset, which is not derived as a point transformation or a coordinate transformation.)
These point transformations seem to represent the action on field configurations taking values in a Hilbert space. The transformations of quantum fields under the Poincar\'{e} group was examined by Wigner\cite{Wigner}, who identified the unitary representations of the group, including vector representations.

It appears there could be merit in a closer study of the relation between coordinate transformations and point transformations, as this could shed light on the relation between the classical/Einsteinian and quantum perspectives.

\section{Conclusions \label{Concs}}

This paper has shown the valuable insights that can be gained from linking tangent space symmetries back to coordinate transformations on the base manifold. These transformations act on the tangent space through elements of the group $GL(4,R)$. We have seen how the coset space decomposition of this group with respect to its Lorentz subgroup helps one to understand the nature of the Lorentz gauge symmetry inherent in teleparallelism. Essentially, the Weitzenb\"{o}ck connection can be constructed using an infinite number of different parallelisms, all related by local Lorentz transformations, so the Lorentz gauge freedom in teleparallelism is a reflection of this freedom of choice of parallelism. 

A change of parallelism can be represented by a Lorentz transformation, which acts on the matrix relating the coordinate basis to the frame basis from the right. However, in general, a change of coordinates will have a Lorentz component to its action on the coordinate basis, and some changes of coordinate can be represented purely as Lorentz transformations. These act on the matrix relating the coordinate basis to the frame basis from the left. These are not independent of the changes of parallelism - the action from the left `feeds through' to $i_0$, inducing a gauge transformation of the Lorentz connection. This means that a choice of Lorentz gauge is not respected under changes of coordinate.

This paper also makes it clear that frame bases can also be coordinate bases along a geodesic, when Riemann normal coordinates are employed. This choice of coordinates provides further insights into the relations between different connections.

Finally, we noted that the homomorphic mapping from the covariance group into the general linear group includes a mapping of the Poincar\'{e} group. For global transformations, translations are in the kernel of this map. Local translations, on the other hand, provide a complete cover of the general linear group, through the derivatives of their parameters. Thus general linear transformations on a given tangent space may be represented as local translations and \emph{vice versa}. However, general linear maps between coordinate and frame basis fields across a coordinate neighbourhood can only be represented as local translations on a Riemann flat spacetime.
Inhomogeneous transformations of the vector components cannot be represented in this way, but can result from a `point translation', as shown in Section \ref{point}.

While we have applied this theory in the case of four-dimensional spacetime, all of it is valid any any number of dimensions, regardless of the action. We assume only that the spacetime is pseudo-Riemannian, that all transformations on the tangent bundle are induced by changes of coordinate, and that observations are made by a classical, point-like observer.

\section*{Acknowledgments}

The author would like to thank Eric Huguet, Morgan Le Delliou, Mar\'{i}a Jos\'{e} Guzm\'{a}n, Tomi Koivisto and Martin Kr\u{s}\u{s}ak for thought-provoking conversations and comments on the topics in this paper. I would also like to thank all those who have translated the classic papers listed below, those who put helpful postings on sites such as StackExchange and ResearchGate and those who have edited Wikipedia pages on these topics. Particular thanks to Alexander Unzicker for the key insights I gained from material on his website.

\end{document}